\documentclass[preprint,showpacs,preprintnumbers,amsmath,amssymb]{revtex4}

\usepackage{bm}% bold math
\usepackage{amsmath,amssymb}
\catcode`@=11
\newcommand\eqalign[1]{\null\,\vcenter{\openup\jot\m@th
\ialign{\strut\hfil$\displaystyle{##}$&$\displaystyle{{}##}$\hfil
\crcr#1\crcr}}\,}
\newcommand\meqalign[1]{\null\,\vcenter{\openup\jot\m@th
\ialign{\strut\hfil$\displaystyle{##}$&&$\displaystyle{{}##}$\hfil
\crcr#1\crcr}}\,}
\newcommand\sgn{\mathop{\rm sgn}\nolimits}

\begin{document}
\title{(1+1)-dimensional separation of variables}
\author{Giuseppe Pucacco}
\affiliation{Dipartimento di Fisica -- Universit\`a di Roma ``Tor Vergata" \\ 
INFN -- Sezione di Roma II}
\email{pucacco@roma2.infn.it} 
\author{Kjell Rosquist} 
\affiliation{Department of Physics -- Stockholm University}
\email{kr@physto.se}

\begin{abstract}
In this paper we explore general conditions which guarantee that the geodesic flow on a 2-dimensional manifold with indefinite signature is locally separable. This is equivalent to showing that a 2-dimensional natural Hamiltonian system on the hyperbolic plane possesses a second integral of motion which is a quadratic polynomial in the momenta associated with a 2nd-rank Killing tensor. We examine the possibility that the integral is preserved by the Hamiltonian flow on a given energy hypersurface only (weak integrability) and derive the additional requirement necessary to have conservation at arbitrary values of the Hamiltonian (strong integrability). Using null coordinates, we show that the leading-order coefficients of the invariant are arbitrary functions of one variable in the case of weak integrability. These functions are quadratic polynomials in the coordinates in the case of strong integrability. We show that for $(1+1)$-dimensional systems there are three possible types of conformal Killing tensors, and therefore, three distinct separability structures in contrast to the single standard Hamilton-Jacobi type separation in the positive definite case. One of the new separability structures is the complex/harmonic type which is characterized by complex separation variables. The other new type is the linear/null separation which occurs when the conformal Killing tensor has a null eigenvector.
\end{abstract}

%%%%%%%%%%%%%%%%%%%%%%%%%%%%%%%%%%%%%%%%%%%%%%%%%%%%%%%%%%%%%%%%%%%%%%%%%%%%%%
%\begin{document}
\maketitle

%\vfill\eject

%%%%%%%%%%%%%%%%%%%%%%%%%%%%%%%%%%%%%%%%%%%%%%%%%%%%%%%%%%%%%%%%%%%%%%%%%%%%%%
\section{Introduction}
The cases of weak and strong separation of the equations of motion on a 2-dimensional positive definite manifold were previously treated by us \cite{geom}. In that work we related separating coordinate systems with 2nd rank Killing tensors based on the approach used by Rosquist and Uggla \cite{ru:kt} where the corresponding problem in the indefinite $(1+1)$-dimensional case was treated with some applications to cosmological space-times. 

In the present work, we make a systematic analysis of the separation of the geodesic equations on 2-dimensional pseudo-Riemannian manifolds classifying all separating coordinate systems. The most important difference with respect to the positive definite case concerns the structure of the traceless conformal Killing tensor. In the positive definite case this structure is unique and leads to a single ``standard'' form. In the indefinite $(1+1)$ case on the other hand, the possible standard forms turn out to be three, leading to a much richer structure. The two new separability structures which appear in addition to standard Hamilton-Jacobi separation are the complex/harmonic and linear/null structures. The complex/harmonic structure is characterized by complex separating variables while the linar/null case occurs when the conformal Killing tensor has a null eigenvector.

Systems with indefinite signature have been investigated before in the classical works by Kalnins \cite{kalnins1, kalnins2}. A different approach used by Rastelli \cite{rastelli} was recently re-examined \cite{cdml,dr}. However a complete and thorough understanding of all aspects has been lacking. In particular, the existence of the new separability structures, complex/harmonic and linear/null (apparently first uncovered in the Rosquist and Uggla  paper \cite{ru:kt}), has not been fully recognized and systematized previously. 

The layout of the paper is as follows: in Sect.~2 we recall the link between geodesic motion, natural Hamiltonian flows and suitable coordinates, reviewing generalized symmetries of these systems associated with Killing tensors; in Sect.~3 we study the separability conditions for indefinite 2-dimensional systems; in Sect.~4 we classify the cases of strong separability that coincide with those of free motion on the flat Minkowski plane; the conclusions are discussed in Sect.~5.

%%%%%%%%%%%%%%%%%%%%%%%%%%%%%%%%%%%%%%%%%%%%%%%%%%%%%%%%%%%%%%%%%%%%%%%%%%%%%%
\section{Geometric formulation of (1+1)-dimensional systems}
We consider the general 2-dimensional indefinite dynamical metric written in the manifestly conformally flat form
\begin{equation}\label{metric}
     ds^2 = 2\Gamma(u,x) (-d u^2 + d x^2)
\end{equation}
where $\Gamma(u,x)>0$. Note that any 2-dimensional geometry with indefinite signature can be represented by the canonical form given in \eqref{metric}. In a relativistic context, the variable $u$ is a time coordinate, while $x$ is a spatial coordinate. In addition to describing the geometry on a hyperbolic 2-dimensional manifold, this metric can also be associated with the natural Hamiltonian (via the standard Jacobi-Maupertuis trick \cite{lanczos:mechanics,bcr})
\begin{equation}\label{eq:ham_indef}
   H= \tfrac12(-p_u^2 + p_x^2) + \Phi (u,x)
\end{equation}
if the identification 
\begin{equation}\label{IDCF}
\Gamma = \Phi - {\cal E}
\end{equation}
is made for a given value ${\cal E}$ (the `energy') of the Hamiltonian function. When dealing with conformal transformations, it is convenient to introduce the `null' Hamiltonian, by incorporating ${\cal E}$ 
\begin{equation}\label{eq:ham_zero}
   {\cal H}= H - {\cal E} = \tfrac12(-p_u^2 + p_x^2) + \Gamma \ .
\end{equation}
The condition $\Gamma>0$ implies ${\cal E} <\Phi$. To consider energies in the region ${\cal E} >\Phi$ one should redefine $\Gamma$ by $\Gamma \rightarrow -\Gamma$ and reinterpret the variables by performing the transformation $u \leftrightarrow x$.
 
To study 2-dimensional systems is very helpful to use coordinates
which are null (lightlike) with respect to the dynamical metric. Such
variables are naturally adapted to the action of the conformal group
which plays an essential role for 2-dimensional systems. The case
of a positive definite dynamical metric was discussed in \cite{geom}.
 The null variables for that case are complex conjugates and the conformal
group can be parametrized by one arbitrary analytic function. In the indefinite
(Lorentzian) signature case 
\cite{ru:kt}, the null variables are real and the conformal
group can be parametrized by two arbitrary real functions of one variable.
We  introduce null coordinates according to
\begin{equation}\label{eq:real}\eqalign{
           \zeta &=u+x \cr
         \hat\zeta &=u-x \ .\cr
}\end{equation}
The metric then becomes
\begin{equation}\label{eq:ds2real}
    ds^2 = -2 \Gamma(\zeta,\hat\zeta) \, d \zeta \, d \hat\zeta 
\end{equation}
and the Hamiltonian is 
\begin{equation}\label{eq:ham_null}
   H= -2p_\zeta p_{\hat\zeta} + \Phi (\zeta, \hat\zeta)\ .
\end{equation}
It is notable that this Hamiltonian, written in the null coordinates, is formally of the same form as the positive definite Hamiltonian 
\begin{equation}
H= 2 p_z p_{\bar z} +V (z, \bar z)
\end{equation} 
where 
\begin{equation}\label{eq:pdreal}\eqalign{
           z &=x+iy \cr
         \bar z &=x-iy \cr
}\end{equation}
with $x,y$ being Cartesian coordinates in ${\mathbb{E}}^{2}$ and $V(z, \bar z)$ is the standard potential energy. The metric is given by
\begin{equation} ds^2 = 2 G(z, \bar z) \, d z \, d \bar z \ . 
\end{equation}
In analogy with \eqref{IDCF}, the conformal factor is defined as
\begin{equation}\label{PDCF}
G = E - V 
\end{equation}
where $E$, the given value of the Hamiltonian function, is the physical energy. In the Euclidean case, only the region $V < E$ is relevant implying $G>0$. The very important difference is that $z$ and $\bar z$ are complex conjugates while $\zeta$ and $\hat\zeta$ in \eqref{eq:ham_null} are real variables. This difference turns out to be crucial and leads to a much richer structure in the indefinite case. However, to start with, the geometric formulation of (1+1)-dimensional systems proceeds in much the same way as the positive definite case. 

We assume that, as usual, the simplest type of integrability (and supposedly of separability) is associated with the conservation of a quadratic integral of motion. We therefore seek conditions on $\Gamma$ which guarantee the existence of a second rank Killing tensor. To that end we follow \cite{ru:kt} as closely as possible. As seen above the dynamics can be represented by a conformally flat metric in the form
\begin{equation}\label{gemet}
    ds^2 =  \Gamma(x^{k}) h_{ij} dx^{i} dx^{j} \quad (i,j,k=0,1)
\end{equation}
where $h_{ij}$ is the metric tensor on the Minkowski plane ${\mathbb{M}}^{2}$. However, when doing calculations it is convenient to employ either an orthonormal frame
$\omega^{\hat I}$ using hatted upper case Latin indices ($\hat I,\hat J,\hat
K,\ldots = \hat0, \hat1$) in terms of which the metric is written as
\begin{equation}  ds^2 = -(\omega^{\hat0})^2 + (\omega^{\hat1})^2
\end{equation}
or a null frame $\Omega^I$ using upper case Latin indices
($I,J,K,\ldots = 0,1$) with the metric taking the form
\begin{equation}  ds^2 = -2 \Omega^0 \Omega^1 \ .
\end{equation}
The frame components are given by
\begin{equation}\label{eq:frame}\meqalign{
   \omega^{\hat0} &= (2\Gamma)^{1/2}du \qquad\qquad
         \Omega^0 &= \omega^{\hat0} + \omega^{\hat1} = \Gamma^{1/2}d \zeta \cr
   \omega^{\hat1} &= (2\Gamma)^{1/2}dx \qquad\qquad
         \Omega^1 &= \omega^{\hat0} - \omega^{\hat1} = \Gamma^{1/2}d\hat\zeta
}\end{equation}
and the metric tensor in the two cases respectively is
\begin{equation}
       (h_{\hat I \hat J}) = \biggl(\begin{matrix}  -1 & 0 \\[-5pt]
                                              0 & 1 \,\end{matrix}\biggr)
\end{equation}
and
\begin{equation}
       (h_{I J}) = -\biggl(\,\begin{matrix} 0 & 1 \\[-5pt]
                                          1 & 0 \end{matrix} \,\biggr) \ .
\end{equation}
The geodesic equations for the metric \eqref{eq:ds2real} and the canonical equations for the Hamiltonian \eqref{eq:ham_null} have the same solutions as unparametrized trajectories. The link between the time parameter along the solution orbit of the canonical equations (say $t$) and the parameter $s$ along the corresponding geodesic in the Jacobi-Maupertuis metric is given by the relation
\begin{equation}
ds = 2 \Gamma dt \ .
\end{equation}

%%%%%%%%%%%%%%%%%%%%%%%%%%%%%%%%%%%%%%%%%%%%%%%%%%%%%%%%%%%%%%%%%%%%%%%%%%%%%%
\section{Separability of indefinite 2-dimensional systems}
\subsection{Killing tensors on the hyperbolic plane}
A second rank Killing tensor is a
symmetric tensor $K_{IJ} = K_{(IJ)}$ satisfying the equation
\begin{equation}\label{eq:kte}
     K_{(IJ;K)} = 0 \ .
\end{equation}
The existence of such a tensor is equivalent to the existence of an invariant of the geodesic equations which is a polynomial of the second degree in the momenta
\begin{equation}\label{inv2}
{\cal I}_2 = K^{IJ} p_I p_J \ .
\end{equation}
Killing tensors and second degree invariants are the natural generalizations
of Killing vectors $K^I$ and the corresponding first degree invariants $K^I
p_I$. Any Killing vector gives rise to a Killing tensor $K_{(I} K_{J)}$. Such
a tensor is said to be {\it reducible}. 

An important property of the second rank Killing tensor equations (\ref{eq:kte}) is that they can be decomposed in conformal
(traceless) and trace parts \cite{ksmh:exact}
\begin{equation}\label{eq:kteq}
 \begin{split}
      C_{IJK} &= P_{(IJ;K)} - \frac12 h_{(IJ} P^L{}_{K);L} = 0 \\[3pt]
       K_{;I} &= - P^J{}_{I;J} \cr
 \end{split}
\end{equation}
where the Killing tensor itself is decomposed in a conformal part $P_{IJ}$ and
the trace $K = K^I{}_I$ according to
\begin{equation}
     K_{IJ} = P_{IJ} + \frac{K}2 h_{IJ} \ .
\end{equation}
Referring to the vector $P_I  = P^J{}_{I;J}$ as the ``conformal current'', it
follows from (\ref{eq:kteq}) that, for any given conformal Killing tensor, the
equation for the trace can be solved if the integrability condition
\begin{equation}\label{eq:intcond1}
     P_{[I;J]} = 0
\end{equation}
is satisfied. The procedure to solve the Killing tensor equations is
therefore firstly to solve the conformal Killing tensor equations and then
check if the integrability condition (\ref{eq:intcond1}) for the trace can be
satisfied. In fact it turns out that the conformal equations $C_{IJK} =0$ can easily be
solved, leaving the integrability condition as the remaining equation to study.

We can represent the null frame components of
the conformal part of the Killing tensor by a pair of real functions $\Sigma, \hat \Sigma$ (cf.~\cite{ru:kt}), so that the Killing tensor can be written as
\begin{equation}\label{eq:kt}
       (K_{MN}) = \begin{pmatrix}
                   \hat\Sigma\Gamma &  K/2 \\
                                K/2 &  \Sigma\Gamma \end{pmatrix} \ .
\end{equation}
In this frame, the only nontrivial components of the conformal Killing tensor equations
are \cite{geom,ru:kt}
\begin{equation}\label{eq:confeq}
         C_{000} = \Gamma^{1/2} \hat\Sigma_{,\zeta} = 0 \ ,\qquad\qquad
         C_{111} = \Gamma^{1/2} \Sigma_{,\hat\zeta} = 0 \ .
\end{equation}
It follows that $P_{MN}$ is a conformal Killing tensor precisely if $\Sigma$ {\it is
an arbitrary function of $\zeta$ only} and $\hat\Sigma$ {\it is
an arbitrary  function of $\hat\zeta$ only}
\begin{equation}\label{cktsol}
         \Sigma = \Sigma (\zeta) \ ,\qquad\qquad
     \hat\Sigma = \hat\Sigma (\hat\zeta) \ .
\end{equation} 
Thus the conformal Killing tensor has the form
\begin{equation}\label{CKTID}
       (P_{MN}) = \Gamma \begin{pmatrix} \hat\Sigma (\hat\zeta) & 0 \\
                                0 &  \Sigma (\zeta)  \end{pmatrix} .
\end{equation}
The equations for the trace are
\begin{equation}\label{eq:traceeq}
 \begin{split}
      K_{,\zeta} &= - 2  \Gamma_{,\hat\zeta} \hat \Sigma (\hat\zeta) - \Gamma \hat \Sigma'(\hat\zeta) \\[3pt]
 K_{,\hat\zeta} &= - 2  \Gamma_{,\zeta} \Sigma(\zeta) - \Gamma \Sigma'(\zeta) \ .
 \end{split}
\end{equation}
The integrability condition \eqref{eq:intcond1} can be written in terms of the coordinates in the form
\begin{equation}
K_{,\zeta \hat\zeta} = K_{,\hat\zeta \zeta} \ .
\end{equation}
In the present indefinite case, the integrability condition becomes
\begin{equation}\label{GDE}
    2\Gamma_{,\zeta\zeta} \Sigma(\zeta) - 2\Gamma_{,\hat\zeta \hat\zeta}\hat \Sigma(\hat\zeta)
  + 3\Gamma_{,\zeta}\Sigma'(\zeta) - 3\Gamma_{,\hat\zeta}\hat \Sigma'(\hat\zeta) 
  + \Gamma[\Sigma''(\zeta)-\hat \Sigma''(\hat\zeta)] = 0 \ .
\end{equation}
We will refer to this equation as the generalized Darboux equation. Its counterpart in the positive definite case reduces to the classical Darboux equation when requiring strong integrability (see \cite{configjmp}).

Once equation \eqref{GDE} is solved, the families of Killing tensors of the system are determined and using the contravariant form of the Killing tensor, the existence of conserved quantities of the form \eqref{inv2} 
is established. We remark that the conserved quantity should be interpreted as an integral of motion for the null Hamiltonian \eqref{eq:ham_null} in a suitable time gauge. In fact, the trace of the Killing tensor depends on ${\cal E}$ and the integral \eqref{inv2} is actually a function of the form
\begin{equation}\label{invar}
{\cal I}_2 =  {\cal I}_2 (p_\zeta, p_{\hat\zeta}, \zeta, \hat\zeta; \cal E)\ .
\end{equation} 
To transform it into the ordinary
integral of motion in the physical time gauge, there is a straightforward recipe
consisting in replacing the parameter $\cal E$ appearing in ${\cal I}_2$ with the
corresponding Hamiltonian function. As a consequence, the
physical integral of motion is
\begin{equation}\label{prescription} 
I_{2} = {\cal I}_2 \big\vert_{{\cal E} \rightarrow H (p_\zeta, p_{\hat\zeta}, \zeta, \hat\zeta)} \ .
\end{equation}
In this case with 2 degrees of freedom, Liouville integrability is obtained. As in the usual setting of positive definite geometry, the existence of quadratic conserved quantities gives more than mere integrability, since in all cases it allows the {\it separability} of the equations of motion associated with the Hamiltonian \eqref{eq:ham_null} as described below.
 
 \subsection{Solution of the integrability condition}

To solve the generalized Darboux equation, it is useful to put it in a simpler form by exploiting a suitable coordinate transformation. As described in \cite{geom,ru:kt}, we use a conformal transformation to standardize the frame and coordinate representation of the conformal Killing tensor. To that end we introduce the new null frame
\begin{equation}
\widetilde\Omega^0 =  B\Omega^0 \ ,\qquad\qquad
\widetilde\Omega^1 =  B^{-1}\Omega^1
\end{equation}
and new coordinates $W$ and $\hat W$ by means of the transformation 
\begin{equation}\label{SSC}
\zeta = F(W) \ ,\qquad\qquad \hat\zeta = \hat F(\hat W) \end{equation} 
with inverse
\begin{equation}\label{NNC}
F^{-1}(\zeta) = A(\zeta)=  W \equiv U +X \ ,\qquad 
\hat F^{-1}(\hat\zeta) = \hat A(\hat \zeta) = \hat W \equiv U- X \ .
\end{equation} 
In the following, we will refer to $U$ and $X$ as {\it separating} variables, because, as shown below, separation of the equations of motion occurs in general, even if in a nonstandard fashion. 
Referring to (\ref{eq:frame}), the choice 
\begin{equation}
B= \sqrt{\hat F'(\hat W) / F'( W)} 
\end{equation} 
implies that the new frame $\widetilde\Omega^I$ has the coordinate representation 
\begin{equation}
\widetilde\Omega^0 = \widetilde \Gamma^{1/2} d  W \ ,\qquad\qquad
\widetilde\Omega^1 = \widetilde \Gamma^{1/2} d\hat  W
\end{equation}
where
\begin{equation}
\widetilde \Gamma = F'( W) \hat F'(\hat W) \Gamma
\end{equation} 
is the new metric conformal factor. The metric in the new coordinates has the same form as (\ref{eq:ds2real})
\begin{equation}
ds^2 = -2\widetilde \Gamma ( W, \hat W)d  W d\hat  W \ .
\end{equation} 
It follows that the conformal Killing tensor components in the
new frame are given by
\begin{equation}
 \begin{split}
   \widetilde P_{00} &= B^{-2} P_{00}
                      = [\hat A'(\hat \zeta)]^2  \hat \Sigma (\hat\zeta)
                          \widetilde \Gamma \\[5pt]
   \widetilde P_{11} &= B^2 P_{11} = [A'(\zeta)]^2 \Sigma (\zeta) 
                          \widetilde \Gamma\ .
 \end{split}
\end{equation}
If $\Sigma$ and $ \hat \Sigma $ are both nonzero, we can choose the functions $A$ and $\hat A$ so that
\begin{equation}
[A'(\zeta)]^{-2} =|\Sigma(\zeta)|\end{equation}
and
\begin{equation}
[\hat A'(\hat\zeta)]^{-2} =|\hat \Sigma(\hat\zeta)| \ .\end{equation}
Note that $\det(P_{IJ})$ is invariant under this frame scaling transformation, implying that the relative sign between $P_{00}$ and $P_{11}$ is also invariant. Choosing $P_{11}$ to be nonnegative, we can always bring the conformal Killing tensor to one of three standard forms, namely
\begin{equation}\label{CKTSF}
\widetilde
                  P_{00} = \epsilon \,  \widetilde \Gamma \ ,\qquad
      \widetilde  P_{11} = \widetilde \Gamma \ ,
\end{equation}
where $\epsilon =\pm1$ or $\epsilon=0$, which also includes the case in which one of $\Sigma$ and $\hat\Sigma$ vanishes. The coordinate transformations (\ref{SSC}-\ref{NNC}) are then given by
\begin{equation}
F( W) = \int \sqrt{|\Sigma(\zeta( W))|} d  W \ , \qquad
\hat F(\hat W) = \int \sqrt{|\hat\Sigma(\hat\zeta(\hat W))|} d \hat W
\end{equation}
and
\begin{equation}
 W = \int \frac{d \zeta}{\sqrt{|\Sigma|}} \ , \qquad
\hat W = \int \frac{d \hat\zeta}{\sqrt{|\hat\Sigma|}} \ .
\end{equation}

By contrast, in the positive definite case the determinant of the conformal Killing tensor is always positive. In fact, as shown in \cite{geom}, its form is given by \cite{nota1}
 \begin{equation}\label{CKTPD}
       (P_{MN}) = G \Biggl( \begin{matrix} \bar S (\bar z) & 0 \\[-3pt]
                                      0 &  S (z) \end{matrix} \Biggr)\ ,
\end{equation}
where $G$ is the positive definite conformal factor defined in \eqref{PDCF} and $S(z)$ is an arbitrary holomorphic function. In this case, the conformal transformation 
\begin{equation}
   w = A(z)\ , \qquad [A'(z)]^{-2} =S(z)
\end{equation}
determines the standardization of the frame
 so that 
 \begin{equation}
\widetilde
   P_{11} = \widetilde  P_{22} = \widetilde G \ ,
\end{equation}
where $\widetilde G = |S| G$. It follows that $\det(P_{IJ}) = |S|^{2} G^{2}$. 
On the other hand, from \eqref{CKTID} and \eqref{CKTSF} we see that in the indefinite case $\det(P_{IJ})= \Psi^{2} \Gamma^{2} = \epsilon \,  \widetilde \Gamma^{2} $ where
\begin{equation}
   \Psi (\zeta,\hat\zeta) = [A'(\zeta)\hat A'(\hat\zeta)]^{-1}
                  = |\Sigma(\zeta)\hat\Sigma(\hat\zeta)|^{1/2} 
\end{equation}
and
\begin{equation}
   \epsilon  = \sgn\bigl(\det(P_{IJ})\bigr)=\sgn\bigl(\Sigma(\zeta)\hat\Sigma(\hat\zeta)\bigr) \ ,
\end{equation}
if $\det(P_{IJ})\neq0$ and $\epsilon=0$ if $\det(P_{IJ})=0$. We can then infer that the separated form of an indefinite 2-dimensional system comes in three different flavors depending on the sign of $\det(P_{IJ})$. If  $\det(P_{IJ})>0$, we have the standard {\it Hamilton-Jacobi separation}, whereas the cases $\det(P_{IJ})<0$ and $\det(P_{IJ})=0$ have no counterpart in positive definite systems. 
These three possibilities affect the form of the generalized Darboux equation \eqref{GDE} in the new variables. In fact, the coordinate transformations leading to the standard form (\ref{CKTSF}) of the conformal Killing tensor, provide the three different equations 
\begin{equation}
 \widetilde \Gamma_{, W W} - \epsilon \ \widetilde \Gamma_{,\hat W \hat W}
  = 0\ , \qquad \epsilon = +1, -1, 0 \ .
\end{equation}
We now examine each theses cases in turn.

\subsection{Positive determinant of the conformal Killing tensor}

In the case $\det(P_{IJ})>0$, the separation variables are given by \eqref{NNC} where the transformations are generated by
\begin{equation}
            [A'(\zeta)]^{-2} = |\Sigma(\zeta)| \ ,\qquad
   [\hat A'(\hat\zeta)]^{-2} = |\hat \Sigma(\hat\zeta)| \ .
\end{equation}
                   In the new variables ($W, \hat W$) the generalized Darboux equation \eqref{GDE} becomes
\begin{equation}\label{GDEP}
 \widetilde \Gamma_{, W W} - \widetilde \Gamma_{,\hat W \hat W} = \widetilde \Gamma_{,UX} = 0 
\end{equation}
the solution of which, like in the positive definite case, is
\begin{equation}\label{pdgamma}
\widetilde \Gamma = B_{1}(U) + B_{2}(X) \ , \end{equation}
with $B_{1}$ and $B_{2}$ arbitrary functions of their argument. The null Hamiltonian then takes the explicitly separated form
\begin{equation}\label{hampd}
{\cal H} = -2 p_{ W} p_{\hat W}+\widetilde \Gamma =
\frac12 (- p_{U}{}^2 + p_{X}{}^2)+B_{1}(U) + B_{2}(X) \ .
\end{equation}
The equations for the trace \eqref{eq:traceeq} become
\begin{equation}\label{eq:tracepd}
      K_{,U} = - 2  \widetilde \Gamma_{,U} \ ,\qquad
      K_{,X} = 2 \widetilde \Gamma_{,X} \ .
\end{equation}
Using \eqref{pdgamma}, the solution is
\begin{equation}
K = 2\left[B_{2}(X) - B_{1}(U) \right]
\end{equation}
      and the second integral of motion \eqref{inv2} can be written as
    \begin{equation}\label{inv2pd}
{\cal I}_2 = p_{ W}^2 + p_{\hat W}^2 + \frac12 K =
\frac12 (p_{X}^2 + p_{U}^2) + B_{2}(X) - B_{1}(U) \ .
\end{equation}  

\subsection{Negative determinant of the conformal Killing tensor}

We now turn to the case $\det(P_{IJ})<0$, which has no counterpart for positive definite systems. The separation variables \eqref{NNC} are given by 
\begin{equation}
          [A'(\zeta)]^{-2} = |\Sigma(\zeta)| \ , \qquad
 [\hat A'(\hat\zeta)]^{-2} = - |\hat \Sigma(\hat\zeta)| \ .
\end{equation}
The form of the generalized Darboux equation written in the separation variables then changes from the wave equation to the Laplace equation
\begin{equation}\label{GDEN}
     \widetilde \Gamma_{, W W} + \widetilde \Gamma_{,\hat W \hat W}
   = \tfrac12 (\widetilde \Gamma_{,UU} + \widetilde \Gamma_{,XX}) = 0 \ .
\end{equation}
It follows that the general solution is a harmonic function given by
\begin{equation}\label{ndgamma}
   \widetilde \Gamma = \Re\{ Q(Z) \} \ ,
\end{equation}
where $Q(Z)$ is an arbitrary holomorphic function of $Z=X+iU$. This means that the system separates if it is written in the complex variables $Z$ and $\bar Z$, since the null Hamiltonian can be written as
\begin{equation}\label{hamnd}
   {\cal H} = p_{Z}^2 + p_{\bar Z}^2
       + \frac12 \left[Q(Z) + \bar Q(\bar Z)\right]\ .
\end{equation}
We therefore refer to this case as {\it harmonic} or {\it complex separation} in contrast to the additive Hamilton-Jacobi separation. The equations for the trace \eqref{eq:traceeq} are
\begin{equation}\label{eq:tracend}
      K_{,Z} + K_{,\bar Z} = 2 i( \widetilde \Gamma_{,\bar Z}- \widetilde \Gamma_{,Z}) \ ,\qquad
      K_{,Z} - K_{,\bar Z} = -2 i(\widetilde \Gamma_{,\bar Z}+ \widetilde \Gamma_{,Z}) \ .
\end{equation}
Using \eqref{ndgamma}, we find that the solution is given by
\begin{equation}
   K = i \left(Q(Z) - \bar Q(\bar Z)\right) \ .
\end{equation}
The second invariant then takes the form
\begin{equation}\label{inv2nd}
{\cal I}_2 = p_{X} p_{U} + \frac12 K = 
i ( p_{Z}^2 - p_{\bar Z}^2) + \frac{i}{2} \left(Q(Z) - \bar Q(\bar Z)\right).
\end{equation} 

\subsection{Vanishing determinant of the conformal Killing tensor}

We finally consider the third type of separation which occurs when $\det(P_{IJ})=0$ and for which there is again no counterpart in the positive definite case. We may then assume $\Sigma\neq0$ and $\hat \Sigma=0$ (or vice versa). The generalized Darboux equation becomes 
\begin{equation}\label{GDEZ}
\widetilde \Gamma_{, W W}=0.
\end{equation} 
One separation variable is $ W=A(\zeta)$ where $A(\zeta)$ satisfies $[A'(\zeta)]^{-2}=|\Sigma(\zeta)|$ as in the previous cases. There is no restriction on the other variable which can therefore be any function independent of $ W$. It follows that the general solution must have the form
\begin{equation}
   \widetilde \Gamma = C(\hat W) W + D(\hat W) \ ,
\end{equation}
where $C(\hat W)$ and $D(\hat W)$ are arbitrary functions. This case is usually referred to as {\it linear} or {\it null separation}. The latter terminology stems from the fact that the Killing tensor has a double null (or lightlike) eigenvector. We can explicitly check the separation observing that the Hamiltonian is
\begin{equation}\label{hamnud}
{\cal H} = -2 p_{ W} p_{\hat W}+C(\hat W) W + D(\hat W) 
\end{equation}
and the second invariant is
    \begin{equation}\label{inv2nud}
{\cal I}_2 = p_{ W}^2 + \frac12 K = p_{ W}^2 - \int C(\hat W) d \hat W \ . \end{equation}
The canonical equation 
\begin{equation} \frac{d \hat  W}{d \tau} = - 2 p_{ W} \ , 
\end{equation}
given by the Hamiltonian \eqref{hamnud}, allows one to write the expression \eqref{inv2nud} in the form
\begin{equation}
   \left( \frac{d \hat W}{d \tau} \right)^{2}
     = 4\left[ {\cal I}_2 + \int C(\hat W) d \hat W \right] \ .
\end{equation}
The other equation of motion can be written in the form
\begin{equation}
\frac{d  W}{d \tau} = \frac{C(\hat W(\tau)) W + D(\hat W(\tau))}
{\sqrt{{\cal I}_2+\int C(\hat W(\tau)) d \hat W}} \ ,
\end{equation}
where we have used the fact that \eqref{hamnud} is a null Hamiltonian to eliminate $p_{\hat W}$. This is a first order ordinary differential equation and it can therefore always be solved, at least up to quadratures.

%\vfill\eject

%%%%%%%%%%%%%%%%%%%%%%%%%%%%%%%%%%%%%%%%%%%%%%%%%%%%%%%%%%%%%%%%%%%%%%%%%%%%%%
\section{Strong separability}
To summarize, in the (1+1)-dimensional case there are two additional types of separation, the harmonic and the linear separations. For all three separation types, one can choose the coordinate transformation factors $\Sigma(\zeta)$ and $\hat\Sigma(\hat\zeta)$ independently of each other, giving rise to several sets of separation variables without counterparts in the positive definite case.

We now focus on the case of {\it strong} integrability which in the present case of a quadratic second invariant actually amounts to strong separability. Considering the definition \eqref{IDCF}, we recall that, by strong separability, we mean that the system is separable for arbitrary values of the energy ${\cal E}$. This is the ordinary separation condition which has as a particular case the free (geodesic) motion on the `flat' ($\Gamma=1$) hyperbolic plane \cite{bcr}. The case in which the system is separable (or integrable) only at certain fixed values of ${\cal E}$ is usually termed {\it weak} separability (or integrability in the general case of nonquadratic invariants). A discussion of this issue in the setting of positive definite systems has recently been given in \cite{configjmp}.

The condition for strong separability is obtained by requiring that the general integrability condition \eqref{GDE} should not depend on the energy ${\cal E}$ leading to
\begin{equation}\label{ICS}
\Sigma''(\zeta)=\hat \Sigma''(\hat \zeta) \ . \end{equation}
We recall that, in the positive definite case, the corresponding condition is \cite{geom}
\begin{equation}
S''(z)=\bar S''(\bar z) \ , \end{equation}
implying that the conformal function has the form 
\begin{equation}
S(z) = az^{2} + \beta z + \gamma \ , 
\end{equation}
where $a$ is a real constant whereas $\beta$ and $\gamma$ are complex. In the present indefinite case, the solutions of \eqref{ICS} are 
\begin{equation}\label{sigmas}
   \Sigma(\zeta) = k \zeta^2 + b \zeta + c \ ,\qquad
   \hat \Sigma(\hat\zeta) = k \hat\zeta^2 + \hat b \hat\zeta + \hat c \ 
\end{equation}
where all arbitrary constants are real. In both cases the total number of free constants is five and the leading order coefficients can be assumed to take the values 1 or 0. 
In the positive definite case, there arises four distinct cases when evaluating the integral
\begin{equation}
w=\int{S(z)^{-1/2}dz} \ ,
\end{equation}
to obtain the separating variables $(w, \bar w)$. This corresponds precisely to the four classical cases of separability \cite{geom}. In the indefinite case, the equations to be integrated are
\begin{equation}
   \frac{dA}{d \zeta} = \frac{1}{\sqrt{|\Sigma(\zeta)|}} \ ,\qquad 
   \frac{d \hat A}{d \hat\zeta} = \frac{1}{\sqrt{|\hat \Sigma(\hat\zeta)|}}\ ,
\end{equation}
where the functions $A(\zeta)$ and $ \hat A(\hat \zeta)$ may assume five distinct forms. The forms are enumerated in table \ref{tab:sep_forms}, where, for the cases with $k=0$, the standard forms with either $c=1$,~$b=0$ or $c=0$,~$b=4$ are chosen and, for the cases with $k\ne0$, 
\begin{equation}
  D = b^2-4kc \ , \quad  \Delta = \tfrac12\sqrt{|D/k|} \ .
\end{equation} 
The corresponding ``hatted'' quantities $ \hat A(\hat \zeta)$, $ \hat F(\hat \zeta)$, $\hat D$ and $\hat \Delta$ are defined in an analogous way in terms of $\hat b$ and $\hat c$. 

The reason for the fifth class of coordinate transformations is due to the fact that, in the case of real variables, the choice of the hyperbolic sine rather than the hyperbolic cosine provides two independent coordinate systems. In the positive definite case, the analogous transformation corresponds to a hyperbolic function of a {\it complex} variable which generates elliptic-hyperbolic coordinates: the choice of the hyperbolic sine rather than the cosine corresponds to a $\pi / 2$ rotation of the foci of the confocal families of coordinate curves.
\begin{table}
  \centering 
    \begin{tabular}{l|l|l|l|l}
      \hline
% after \\ : \hline or \cline{col1-col2} \cline{col3-col4} ...
 $1.$ & $F_1( W) =  W$ &      $A_1(\zeta) = \zeta$ &  $\Sigma_1(\zeta) = 1$ & 
         									$k=b=0$\ ,\ $c\neq0$ \\
 $2.$ & $F_2( W) =  W^2$  &   $A_2(\zeta) = \sqrt\zeta$ &  $\Sigma_2(\zeta) = 4\zeta$ &
         									$k=c=0$\ ,\ $b\neq0$ \\
 $3.$ & $F_3( W) = e^W$  &   $A_3(\zeta) = \ln\zeta$ &  $\Sigma_3(\zeta) = \zeta^2$ &
       							  		$k\neq0$\ ,\ $D=0$ \\
 $4.$ & $F_4( W) = \Delta\cosh W$  &  $A_4(\zeta) = {\rm arccosh}(\zeta/\Delta)$ &
           $\Sigma_4(\zeta) = \zeta^2-\Delta^2$ &  $k\neq0$\ ,\ $D>0$ \\
 $5.$ & $F_5( W) = \Delta\sinh W$  &  $A_5(\zeta) ={\rm arcsinh}(\zeta/\Delta)$ &
           $\Sigma_5(\zeta) = \zeta^2+\Delta^2$ &  $k\neq0$\ ,\ $D<0$ \\
\hline
\end{tabular}
  \caption{The possible conformal transformation functions for (1+1)-dimensional integrable Hamiltonians with a second degree invariant.}\label{tab:sep_forms}
\end{table}
When combining the five cases, we must consider only cases with the same value of the separating constant $k$, since $k$ appears both in $\Sigma$ and $\hat \Sigma$. There are no other restrictions so this gives four cases with $k=0$ and nine cases with $k\neq0$, thirteen cases in total.
However, it is reasonable not to distinguish systems which can be transformed into each other by the transformation $(\zeta,\hat\zeta) \rightarrow (\hat\zeta,\zeta)$ or equivalently $x \rightarrow -x$. This reduces the number of cases to three for $k=0$ and six for $k\neq0$, nine cases in total. Using the numbers $1$--$5$ appearing in the first column of the table and the corresponding ``hatted'' figures $\hat 1$--$\hat 5$, in an obvious notation, the set of possible independent separating coordinates is given by the combinations
\begin{equation}
 \begin{aligned}
 k &= 0:  \quad 1 \hat 1 \;\;\; 1 \hat 2 \;\;\; 2 \hat 2 \\
 k &\ne 0:  \quad 3 \hat 3 \;\;\; 3 \hat 4\;\;\; 3 \hat 5 \;\;\; 4 \hat 4 \;\;\; 4 \hat 5 \;\;\; 5 \hat 5 \ .
 \end{aligned}
\end{equation}
For each of these possible combinations of the functions $\Sigma$ and $\hat \Sigma$, we need to check the possible values of $\epsilon$ ($\pm1$ or $0$) depending on the sign of $\det(P_{IJ})$ [cf.~\eqref{CKTSF}]. It is straight-forward to establish that the negative sign can only be present when $\Sigma_1, \Sigma_2$ and $\Sigma_4$ are involved, whereas $\det(P_{IJ})$ can only vanish when $\Sigma_1$ and $\Sigma_2$ are involved. 

Let us now consider the case of standard Hamilton-Jacobi separation. The separating coordinates and conformal factors are enumerated in table \ref{tab:conf_sep}. In each of these coordinate systems, we have solutions with the metric factor given by $\widetilde \Gamma = \Psi \Gamma$, where 
\begin{equation}
\Psi = \frac{1}{A'(\zeta) \hat A'(\hat\zeta)} = F'( W) \hat F'(\hat W) \ . \end{equation} 
We recall from \eqref{IDCF} that the true potential is given by
\begin{equation}\label{SPP}
\Phi = \Gamma +  {\cal E} = \frac{\widetilde \Gamma}{\Psi} +  {\cal E} \ . \end{equation}
In the case of strong integrability, it turns out that for every possible coordinate system of table \ref{tab:conf_sep}, the conformal factor associated with the corresponding coordinate transformation separates in the form
\begin{equation}
\Psi=\Psi_{1}(U)+\Psi_{2}(X) \ .
\end{equation}
Using \eqref{SPP}, we can then write the physical potential in the form
\begin{equation}\label{SEPP}
\Phi = \frac{f_{1}(U)+f_{2}(X)}{\Psi_{1}(U)+\Psi_{2}(X)} \ ,
\end{equation} 
where $f_{1}$ and $f_{2}$ are arbitrary functions \cite{nota2}. Referring to \eqref{hampd} we have
\begin{equation}
 \begin{split}
B_{1}(U) &= f_{1}(U)-  {\cal E} \Psi_{1}(U) \\[3pt]
B_{2}(X) &= f_{2}(X)-  {\cal E} \Psi_{2}(X)
 \end{split}
\end{equation}
so that the null Hamiltonian becomes
\begin{equation}\label{Hsep}
{\cal H} =\frac12 (- p_{U}{}^2 + p_{X}{}^2)+f_{1}(U) + f_{2}(X) - 
{\cal E} (\Psi_{1}(U) + \Psi_{2}(X)) \ .
\end{equation}
The physical Hamiltonian in the separating coordinates is therefore given by
\begin{equation}\label{Hp}
H =\frac{1}{\Psi_{1}(U)+\Psi_{2}(X)}
\bigg[\frac12 (- p_{U}{}^2 + p_{X}{}^2)+f_{1}(U) + f_{2}(X)\bigg]\ .
\end{equation}
and, applying the prescription \eqref{prescription}, the second integral of motion is
\begin{equation}\label{invsep}
I_2 = \frac{1}{\Psi_{1}(U)+\Psi_{2}(X)}
\bigg[\Psi_{2}(X)(p_{U}^2 - 2 f_{1}(U)) + \Psi_{1}(U) (p_{X}^2 + 2 f_{2}(X)) \bigg] \ .
\end{equation}  

Turning now to the case of harmonic separation, the conformal factor always separates in the form
\begin{equation}
\Psi=\tfrac12 \bigl(\psi(Z)+\bar\psi(\bar Z) \bigr) = \Re\{\psi(Z) \}\ ,
\end{equation}
as shown below. Proceeding as in the Hamilton-Jacobi case above, we can write the physical potential in the form
\begin{equation}\label{SEPPnd}
\Phi = \frac{Q_{1}(Z)+\bar Q_{1}(\bar Z)}{\psi(Z)+\bar\psi(\bar Z)} =
\frac{\Re\{Q_{1}(Z) \}}{\Re\{\psi(Z) \}}\ ,
\end{equation} 
where $Q_{1}(Z)$ is an arbitrary function. Referring now to \eqref{hamnd} we can write
\begin{equation}
Q(Z) =Q_{1}(Z) -  {\cal E} \psi(Z) \ .
\end{equation}
The leads to the null Hamiltonian
\begin{equation}\label{Hsepnd}
{\cal H} = \Re\{2 p_{Z}^2 + Q_{1}(Z) -  {\cal E} \psi(Z)\} \ .\end{equation}
The physical Hamiltonian in separating coordinates is therefore found to be
\begin{equation}\label{Hpnd}
H =\frac{\Re\{2 p_{Z}^2 + Q_{1}(Z) \}}{\Re\{\psi(Z) \}} \ ,
\end{equation}
while the second integral of motion becomes
 \begin{equation}\label{invsepnd}
I_2 =  \frac{\Re\{-i \bar\psi(2 p_{Z}^2 + Q_{1}(Z)) \}}{\Re\{\psi(Z) \}} \ .
\end{equation}  

Finally considering linear separation, the simplest case is that in which $\Sigma = \Sigma_1 =1$, $\hat\Sigma = 0$ (Cartesian-zero case). The Hamiltonian and the second integral are then given by (\ref{hamnud}) and (\ref{inv2nud}) and are therefore already in the desired separated form. In the case $\Sigma = \Sigma_2 = 4\zeta$, $\hat\Sigma = 0$ (parabolic-zero case), the Hamiltonian and the second integral are expressed in separating coordinates as
\begin{equation}\label{Hpz}
H =\frac{1}{2(U+X)}
\bigg[\frac12 (- p_{U}{}^2 + p_{X}{}^2)+(U+X) Y'(U-X) + D(U-X)\bigg]\end{equation}
and
 \begin{equation}\label{invsepz}
I_2 = \frac{1}{4(U+X)} (p_{U}+p_{X})
       \bigl[(3X-U)p_{U} + (3U-X)p_{X}+4(U-X) D \bigr]  - Y + (U-X) Y' \ ,
\end{equation}  
where $Y$ and $D$ are arbitrary functions of $\hat W = U-X$.

\begin{table}
  \centering 
    \begin{tabular}{l|l|l|l|l}
      \hline
% after \\ : \hline or \cline{col1-col2} \cline{col3-col4} ...
 $1 \hat 1.$ & $F_1( W) =  W$ & $\hat F_{1}(\hat W) = \hat W$ &  $\Psi_{1} = 1$ & $ \Psi_{2} = 1 $ \\
 $1 \hat 2.$ & $F_1( W) =  W$ & $\hat F_{2}(\hat W) = \hat W^{2}$ &  $\Psi_{1}  = 2 U$ &
         									$ \Psi_{2} = - 2 X$ \\
 $2 \hat 2.$ & $F_2( W) =  W^{2}$ & $\hat F_{2}(\hat W) = \hat W^{2}$ &  $\Psi_{1}  = 4 U^{2}$ &
         									$ \Psi_{2} = - 4 X^{2}$ \\
 $3 \hat 3.$ & $F_3( W) = e^W$  &   $\hat F_{3}(\hat W) = e^{\hat W}$ &  $\Psi_{1}  = e^{2 U}$ &
       							  		$\Psi_{2}=0$ \\
 $3 \hat 4.$ & $F_3( W) = e^W$  &  $\hat F_4(\hat W) = \Delta\cosh{\hat W}$ &
           $\Psi_{1}  = \frac12 \Delta e^{2 U}$ &  $\Psi_{2}= - \frac12 \Delta e^{2 X}$ \\
 $3 \hat 5.$ & $F_3( W) = e^ W$  &  $\hat F_5(\hat W) = \Delta\sinh{\hat W}$ &
           $\Psi_{1}  = \frac12 \Delta e^{2 U}$ &  $\Psi_{2}= \frac12 \Delta e^{2 X}$ \\
           $4 \hat 4.$ & $F_4( W) = \Delta\cosh{ W}$  &  $\hat F_4(\hat W) = \Delta\cosh{\hat W}$ &
           $\Psi_{1}  = \frac12 \Delta^{2} \cosh{U}$ &  $\Psi_{2}= - \frac12 \Delta^{2} \cosh{X}$ \\
           $4 \hat 5.$ & $F_4( W) = \Delta\cosh{ W}$  &  $\hat F_5(\hat W) = \Delta\sinh{\hat W}$ &
           $\Psi_{1}  = \frac12 \Delta^{2} \sinh{U}$ &  $\Psi_{2}= \frac12 \Delta^{2} \sinh{X}$ \\
 $5 \hat 5.$ & $F_5( W) = \Delta\sinh{ W}$  &  $\hat F_5(\hat W) = \Delta\sinh{\hat W}$ &
           $\Psi_{1}  = \frac12 \Delta^{2} \cosh{U}$ &  $\Psi_{2}= \frac12 \Delta^{2} \cosh{X}$ \\
\hline
\end{tabular}
  \caption{The possible conformal factors for separable (1+1)-dimensional Hamiltonians.}\label{tab:conf_sep}
\end{table}

In the rest of the section we list all possible coordinate systems and the corresponding allowed Hamiltonian and integral of motion as classified by the sign of the determinant of the conformal Killing tensor. Each item of the list is therefore labelled by the symbol $ (n \hat n)_{\epsilon} $, where $n$ takes the values $1,2,3,4,5$ according to the combinations above and the index $\epsilon$ takes the values $+,-,0$ for the types with respectively positive, negative or zero determinant. When $\epsilon=-1$ or $\epsilon=0$, the coordinates and conformal factors of table \ref{tab:conf_sep} do not apply and are instead given by the forms listed below. Referring to table \ref{tab:sep_forms}, the choice of the parameters in (\ref{sigmas}) are standardized by suitable transformations to give the simplest expressions. In particular we always assume $\Delta=1$ and $\hat\Delta =1$.

%----------------------------------------------------------------------------%
\subsection{The Cartesian-Cartesian ($1\hat1$) case}

%The $1 \hat 1$ case is trivial. All types $1 \hat 1$:P, $1 \hat 1$:N and $1 
%\hat 1$:Z are admitted.\\

In this case the system separates in the physical null coordinates which coincide with the separating coordinates. Actually, in order to simplify expressions, we do not include dilations and rotations, which can be introduced by assuming $c \ne \hat c \ne 1$ in (\ref{sigmas}). The conformal functions and coordinate transformations have the trivial forms
\begin{equation}
 \begin{split}
   &\Sigma_1(\zeta) = 1 \ ,\quad \hat\Sigma_1(\hat\zeta) = 1 \\[3pt]
   &F_1( W) =  W=\zeta = A_1(\zeta) \\
   &\hat F_1(\hat W) = \hat W = \epsilon \hat\zeta = \hat A_1(\hat\zeta) \ ,\quad  \epsilon =
    \begin{cases}
            +1 & \text{Hamilton-Jacobi separation} \\
            -1 & \text{complex/harmonic separation} \\
       \hfill0 & \text{linear/null separation}
    \end{cases}  \\[3pt]
   &U = u \ ,\quad X = x \ , \quad  \epsilon =1 \\[1pt]
   &U = x \ ,\quad X = u \ , \quad  \epsilon =-1\ .
 \end{split}
\end{equation}
In the linear/null case, only one separating variable is actually fixed, but it is convenient to use $\hat W (= \hat\zeta)$ as the second variable.

%----------------------------------------------------------------------------%
\subsubsection{Case $(1\hat1)_+$}
The Hamiltonian and the second invariant are given by
\begin{equation}
            H = \tfrac12(-p_u{}^2+p_x{}^2) + f_1(u) + f_2(x) \ ,\qquad
   \tilde I_2 = \tfrac12 p_x{}^2 + f_2(x) \ .
\end{equation}

%----------------------------------------------------------------------------%
\subsubsection{Case $(1\hat1)_-$}
Defining the complex variable $z= x+iu$, the Hamiltonian and the second invariant are given by
\begin{equation}
 \begin{split}
    H &= \tfrac12(-p_u{}^2+p_x{}^2) + \Re\{Q(z)\}
       = p_z{}^2 + p_{\bar z}{}^2 - \tfrac12[Q(z)+\bar Q(\bar z)] \\[7pt]
  I_2 &= p_u p_x - \Im\{Q(x+iu)\}\ ,\qquad \tilde I_2 = p_z{}^2 + \tfrac12Q(z)\ ,
 \end{split}
\end{equation}
where $Q(z)$ is an arbitrary complex function.

%----------------------------------------------------------------------------%
\subsubsection{Case $(1\hat1)_0$}
In this case there is just one separating variable which we take as $ W= \zeta = u+x$. It is convenient to use the complementary null variable $\hat W= \hat\zeta= u-x$ as the other independent variable. The Hamiltonian and the second invariant are then given by the expressions
\begin{equation}
 \begin{split}
     H &= \tfrac12(-p_u{}^2 + p_x{}^2) + (u+x)Y'(u-x) + D(u-x)
        = -2p_\zeta p_{\hat\zeta} + \zeta\, Y'(\hat\zeta) + D(\hat\zeta) \ ,\\
   I_2 &= \tfrac14(p_u+p_x)^2 - Y(u-x)
        = p_\zeta{}^2 - Y(\hat\zeta) \ ,
 \end{split}
\end{equation}
where $Y$ and $D$ are arbitrary functions.

%----------------------------------------------------------------------------%
\subsection{The Cartesian-parabolic ($1\hat2$) case}

This is the first nontrivial case. It combines rotated Cartesian and parabolic coordinates. As in the Cartesian-Cartesian case we do not include dilations and rotations. The conformal coordinate transformation is given by
\begin{equation}\label{conformal_trafo}
 \begin{split}
 &\Sigma_1(\zeta) = 1 \ ,\quad \hat\Sigma_2(\hat\zeta) = 4\hat\zeta \\[3pt]
 & W = A_1(\zeta) = \zeta \ , \quad F_1( W) =  W \\
 & \hat W = \hat A_2(\hat\zeta) = \sqrt{\epsilon \hat\zeta} \ ,\quad \hat F_2(\hat W) = \epsilon \hat W^{2}   \ ,
 \quad  \epsilon = \begin{cases}
            +1 & \text{Hamilton-Jacobi separation} \\
            -1 & \text{complex/harmonic separation} \\
       \hfill0 & \text{linear/null separation}
    \end{cases}  \\[7pt]
   &U =  \tfrac12(u+x + \sqrt{u-x}) \ ,\quad
   X = \tfrac12(u+x - \sqrt{u-x}) \ ,\quad  \epsilon = +1 \ , \quad u>x\\[4pt]
&U =  \tfrac12(u+x + \sqrt{x-u}) \ ,\quad
   X = \tfrac12(u+x - \sqrt{x-u}) \ ,\quad  \epsilon = -1 \ , \quad u<x \ .
   \end{split}
\end{equation}
Again, in the linear/null case, only one separating variable is actually fixed, but it is convenient to use $\hat W (= \hat\zeta)$ as the second variable.
   
%----------------------------------------------------------------------------%
\subsubsection{Case $(1\hat2)_+$}
The conformal factor is $\Psi = 2 \hat W = 2(U-X) = 2 \sqrt{u-x}$, so that
\begin{equation}
\Psi_{1} = 2 U \ , \quad \Psi_{2} = -2 X \ .
\end{equation}
This leads to the physical potential
\begin{equation}\label{SEPP12+}
\Phi = \frac{f_{1}(U)+f_{2}(X)}{U - X} \ .
\end{equation} 
The Hamiltonian and the second invariant are given by
\begin{equation}
   H =  \tfrac12(-p_u{}^2+p_x{}^2)
      + \frac{f_1\bigl(u+x + \sqrt{u-x}\,\bigr)
      + f_2\bigl(u+x - \sqrt{u-x}\,\bigr)}{\sqrt{u-x}}
\end{equation}
and
\begin{equation}
I_2 = \tfrac14(p_u+p_x)^2 - 2 (p_u - p_x) (u p_x + x p_u)+ 
         \frac{\bigl(u+x - \sqrt{u-x}\,\bigl) f_1(U)
               + \bigl(u+x + \sqrt{u-x}\,\bigr) f_2(X)} {\sqrt{u-x}} \ .
\end{equation}
Here and below it is understood that the function arguments $U$ and $X$ should be regarded as functions of $u$ and $x$ via the relevant coordinate transformation, in this case given by \eqref{conformal_trafo}.

%----------------------------------------------------------------------------%
\subsubsection{Case $(1\hat2)_-$}
Using the complex variable $Z= X+iU$, the separating variables are
\begin{equation}
Z=\tfrac12\Bigl[(1+i)(u+x) - (1-i)\sqrt{x-u}\Bigr] \ , \quad
\bar Z=\tfrac12\Bigl[(1-i)(u+x) - (1+i)\sqrt{x-u}\,\Bigr]
\end{equation}
so that the conformal factor is
\begin{equation}
\Psi = 2 (X-U) = (1+i) Z + (1-i) \bar Z \ .
\end{equation}
This leads to a physical potential of the form
\begin{equation}\label{SEPP12-}
   \Phi = \frac{\Re\Bigl\{Q_{1}\Bigl(\tfrac12\bigl[(1+i)(u+x)
                     - (1-i)\sqrt{x-u}\,\bigr]\Bigr) \Bigr\}}{2\sqrt{x-u}}
\end{equation} 
where $Q_{1}$ is an arbitrary function. The Hamiltonian and the second invariant are given by
\begin{equation}
   H = \tfrac12(-p_u{}^2+p_x{}^2) + \frac{\Re\{Q_{1}(Z)\}}{2\sqrt{x-u}}
\end{equation}
and
\begin{equation}
I_2 = \tfrac14(p_u+p_x)^2 - 2 (p_u - p_x) (u p_x + x p_u)+
         \frac{\Re\bigl\{\bigl(u + x - i \sqrt{x-u}\,\bigr) Q_{1}(Z)\bigr\}}
              {\sqrt{x-u}} \ .
\end{equation}

%----------------------------------------------------------------------------%
\subsubsection{Case $(1\hat2)_0$}
This subcase coincides for all expressions with the case $(1\hat1)_0$ above.

%----------------------------------------------------------------------------%
\subsection{The parabolic-parabolic $(2 \hat 2)$ case}

In this purely parabolic case the conformal coordinate transformation is given by
\begin{equation}
 \begin{split}
  &\Sigma_2(\zeta) = 4 \zeta ,\quad\hat\Sigma_2(\hat\zeta) = 4\hat\zeta\\[3pt]
  &W = A_2(\zeta) =  \sqrt{\zeta} \ , \quad F_2( W) =  W^2 \\
  &\hat W = \hat A_2(\hat\zeta)
          = \sqrt{\epsilon \hat\zeta} \ ,\quad \hat F_2(\hat W)
          = \epsilon \hat W^{2}  \ ,\quad\epsilon = \begin{cases}
            +1 & \text{Hamilton-Jacobi separation} \\
            -1 & \text{complex/harmonic separation} \\
       \hfill0 & \text{linear/null separation}
    \end{cases} \\[5pt]
       &U =  \tfrac12\bigl( \sqrt{u+x} + \sqrt{u-x}\,\bigr) \ ,\quad
       X = \tfrac12\bigl(\sqrt{ u+x} - \sqrt{u-x}\,\bigr) \ ,\quad
         \epsilon = +1 \ ,\quad u^2-x^2 > 0 \\[3pt]
    &U =  \tfrac12\bigl(\sqrt{ u+x} + \sqrt{x-u}\,\bigr) \ ,\quad
        X = \tfrac12\bigl(\sqrt{u+x} - \sqrt{x-u}\,\bigr) \ ,\quad
          \epsilon = -1 \ , \quad u^2-x^2 < 0  \ .
 \end{split}
\end{equation}
In the linear/null case we assume $W = A_1(\zeta) =  \sqrt{\zeta}$ for the first variable and keep $\hat \zeta (= \hat W)$ as the second variable.

%----------------------------------------------------------------------------%
\subsubsection{Case $(2\hat2)_+$}
The conformal factor is
\begin{equation}
   \Psi = 4  W  \hat  W = 4 (U^2-X^2) = 4 \sqrt{u^2-x^2}
\end{equation}
leading to
\begin{equation}
\Psi_{1} = 4 U^2 \ , \quad \Psi_{2} = -4 X^2 \ .
\end{equation} 
The physical potential then takes the form
\begin{equation}
\label{SEPP22+}
\Phi = \frac{f_{1}(U)+f_{2}(X)}{U^2-X^2} \ .
\end{equation} 
The Hamiltonian and the second invariant are given by
\begin{equation}
   H = \tfrac12(-p_u{}^2+p_x{}^2)
      +\frac{f_1\bigl(\sqrt{u+x}+\sqrt{u-x}\,\bigr)
            +f_2\bigl(\sqrt{u+x}-\sqrt{u-x}\,\bigr)}
            {\sqrt{u^2-x^2}}
\end{equation}
and
\begin{equation}
I_2 = p_x (u p_x + x p_u)+ \frac{\bigl(u - \sqrt{u^2-x^2}\,\bigr) f_1(U)
     + \bigl(u + \sqrt{u^2-x^2}\,\bigr) f_2(X)}{\sqrt{u^2-x^2}} \ .
\end{equation}

%----------------------------------------------------------------------------%
\subsubsection{Case $(2\hat2)_-$}
Using the complex variable $Z= X+iU$, the separating coordinates are
\begin{equation}
Z=\tfrac12(1+i)\bigl(\sqrt{u+x} +i\sqrt{x-u}\,\bigr) \ , \quad
\bar Z=\tfrac12(1-i)\bigl(\sqrt{u+x} -i\sqrt{x-u}\,\bigr)
\end{equation}
corresponding to the conformal factor
\begin{equation}
\Psi = 4 \sqrt{x^2-u^2} = 2 ( Z^2 + {\bar Z}^2) \ .
\end{equation}
This leads the physical potential
\begin{equation}\label{SEPP22-}
   \Phi = \frac{\Re\bigl\{Q_{1}\bigl(\tfrac12(1+i)
                 \bigl[\sqrt{u+x} +i\sqrt{x-u}\,\bigr] \bigr) \bigr\}}
               {\sqrt{x^2-u^2}}
\end{equation} 
where $Q_{1}$ is an arbitrary function. The Hamiltonian and the second invariant are given by
\begin{equation}
   H = \tfrac12(-p_u{}^2+p_x{}^2) + \frac{\Re\{Q_{1}(Z)\}}{\sqrt{x^2-u^2}}
\end{equation}
and
\begin{equation}
I_2 = p_x (u p_x + x p_u) + \frac{\Re\bigl\{\bigl(u - i \sqrt{x^2-u^2}\,\bigr)
   Q_{1}(Z)\bigr\}}{\sqrt{x^2-u^2}} \ .
\end{equation}

%----------------------------------------------------------------------------%
\subsubsection{Case $(2\hat2)_0$}
In this case there is just one separating variable which we take as $W =  \sqrt{\zeta}$. It is convenient to use the complementary null variable $\hat W= \hat\zeta= u-x$ as the other independent variable. The Hamiltonian and the second invariant are then given by
\begin{equation}
 \begin{split}
     H &= \tfrac12(-p_u{}^2 + p_x{}^2) + \sqrt{u+x} Y'(u-x) + D(u-x) \\[3pt]
   I_2 &= (p_u + p_x) (p_x u + p_u x) - Y(u-x)
         + \frac{u-x}{\sqrt{u+x}} D(u-x) + (u-x) Y'(u-x)
 \end{split}
\end{equation}
where $Y$ and $D$ are arbitrary functions.

%----------------------------------------------------------------------------%
\subsection{The polar-polar ($3 \hat 3$) case}

The case $(3 \hat 3)$ corresponds to polar coordinates and admits only standard (Hamilton-Jacobi) separation. The conformal coordinate transformation is
\begin{equation}
 \begin{split}
 &\Sigma_3 (\zeta) = \zeta^{2} \ ,\quad\hat\Sigma_3(\hat\zeta)
                   = \hat\zeta^{2} \\[3pt]
 &  W = A_3(\zeta) =  \ln{|\zeta|} \ , \quad F_3( W) = e^{ W} \\[3pt]
 &  \hat W = \hat A_3(\hat\zeta) = \ln{|\hat\zeta|} \ ,\quad \hat F_3(\hat W) = e^{\hat  W}
 \end{split}
\end{equation}
or in non-null coordinates
\begin{equation}
   U = \frac12\ln(|u^2-x^2|) \ ,\qquad
   X = {\rm arctanh} \left(\frac{x}{u}\right) \ .
\end{equation}
The corresponding conformal factor is 
\begin{equation}
\Psi \equiv \Psi_{1} = e^{W + \hat  W} = e^{2U} = u^2-x^2 \ .
\end{equation} 
This leads to the physical potential
\begin{equation}
\label{SEPP33}
\Phi = \frac{f_{1}(U)+f_{2}(X)}{e^{2U}} \ .
\end{equation} 
The Hamiltonian and the second invariant are given by
\begin{equation}
   H =  \tfrac12(-p_u{}^2+p_x{}^2) + 
   \frac{f_1\bigl(\frac12\ln(|u^2-x^2|)\bigr) + 
           f_2\bigl({\rm arctanh} \bigl(\frac{x}{u}\bigr)\bigr)}{u^2-x^2}
\end{equation}
and
\begin{equation}
I_2 = (u p_x + x p_u)^{2}+ 2 f_2(X) \ .
\end{equation}

%----------------------------------------------------------------------------%
\subsection{The polar-elliptical case of the first kind ($3 \hat 4$)}
The cases $(3 \hat 4)_{\epsilon}$ combine polar and elliptical coordinates and admit standard and harmonic separation. The conformal coordinate transformation is (with $\hat \Delta = 1$)
\begin{equation}
 \begin{split}
 \Sigma_3 (\zeta) &= \zeta^{2} ,\qquad \hat\Sigma_4(\hat\zeta) = \hat\zeta^{2} - 1\ ,\qquad
  W = A_3(\zeta) =  \ln{|\zeta|}  \ , \qquad F_3( W) = e^{ W} \\[5pt]
  \hat W &= \hat A_4(\hat\zeta) = \int \frac{d \hat\zeta}{\sqrt{\epsilon(\hat\zeta^{2}-1)}} = \\[5pt]
   &\begin{cases}
 { \rm arcosh} \!\hat\zeta \ ,&\hat F_4(\hat W)
   = \cosh{\hat W}\ ,\quad\epsilon=+1\ , 
    \quad \text{Hamilton-Jacobi separation} \\[3pt]
  \arcsin{\hat\zeta}\ , & \hat F_4(\hat W) = \sin{\hat  W} \ ,
    \quad \epsilon=-1\ , 
    \quad \text{harmonic separation}    \end{cases}
 \end{split}
\end{equation}
or in non-null coordinates
\begin{equation}
 \left.\begin{split}
   U &= \tfrac12 \Bigl[ \ln(u+x) + \ln(u-x+\sqrt{(u-x)^{2}-1})\Bigr]\\[3pt]
   X &= \tfrac12 \Bigl[ \ln(u+x) - \ln(u-x+\sqrt{(u-x)^{2}-1})\Bigr]
 \end{split} \ \right\} \quad \epsilon=+1 \ , \quad |u-x| > 1 
\end{equation}
and
\begin{equation}
  \left. \begin{split}
   U &= \tfrac12 \Bigl[ \ln(u+x) + \arcsin(u-x)\Bigr] \\[3pt]
   X &= \tfrac12 \Bigl[ \ln(u+x) - \arcsin(u-x)\Bigr] 
 \end{split} \ \right\} \quad \epsilon=-1 \ , \quad |u-x| < 1 \ .
\end{equation}

%----------------------------------------------------------------------------%
\subsubsection{Case $(3\hat4)_+$}
The conformal factor in this standard Hamilton-Jacobi case is given by
\begin{equation}
\Psi = e^{W} \sinh{ \hat  W} = (u+x) \sqrt{(u-x)^{2}-1}\end{equation} 
so that
\begin{equation}
\Psi_{1} = \tfrac12 e^{2 U} \ , \quad \Psi_{2} = -\tfrac12 e^{2 X} \ .\end{equation} 
The physical potential then takes the form
\begin{equation}\label{SEPP34+}
\Phi = \frac{f_{1}(U)+f_{2}(X)}{e^{2U}-e^{2 X}} \ .
\end{equation} 
The Hamiltonian and the second invariant are given by
\begin{equation}
   H =  \tfrac12(-p_u{}^2+p_x{}^2) + 
   \frac{f_{1}(U)+f_{2}(X)}{(u+x) \sqrt{(u-x)^{2}-1}}
\end{equation}
and
%\begin{equation}
%I_2 =  (u p_x + x p_u)^{2} - \tfrac14 (p_u - p_x)^{2}+ 
%  \frac{(u-x-\sqrt{(u-x)^{2}-1}) f_{1}(U) +(u-x+\sqrt{(u-x)^{2}-1}) f_2(X)}
%       {\sqrt{(u-x)^{2}-1}}  \ .
%\end{equation}
%
\begin{equation}
I_2 =  (u p_x + x p_u)^{2} - \tfrac14 (p_u - p_x)^{2} + 
  \left[\frac{u-x}{\sqrt{(u-x)^2-1}} - 1\right]f_1(U) +
  \left[\frac{u-x}{\sqrt{(u-x)^2-1}} + 1\right]f_2(X) \ .
\end{equation}

%----------------------------------------------------------------------------%
\subsubsection{Case $(3\hat4)_-$}
In this harmonic case the separating variables are
\begin{equation}
Z=\tfrac12(1+i)\bigl[\ln(u+x) +i\arcsin(x-u)\bigr] \ , \quad
\bar Z=\tfrac12(1-i)\bigl[\ln(u+x) -i\arcsin(x-u)\bigr]
\end{equation}
so that the conformal factor
\begin{equation}
\Psi = e^{W} \cos{ \hat  W} = (u+x) \sqrt{1 - (u-x)^{2}} 
     = \tfrac12 \bigl[e^{(1-i)Z} + e^{(1+i){\bar Z}}\bigr] \ .
\end{equation}
can be written in the form
\begin{equation}
\Psi = \tfrac12 \bigl[e^{(1-i)Z} + e^{(1+i){\bar Z}}\bigr] \ .
\end{equation}
This leads to a physical potential of the form
\begin{equation}\label{SEPP34-}
\Phi = \frac{\Re\Bigl\{Q_1
  \Bigl(\tfrac12(1+i)\bigl[\ln(u+x) +i\arcsin(x-u)\bigr]\Bigr) \Bigr\}}
            {(u+x) \sqrt{1 - (u-x)^{2}}}
\end{equation} 
where $Q_{1}$ is an arbitrary function. The Hamiltonian and the second invariant are given by
\begin{equation}
   H = \tfrac12(-p_u{}^2+p_x{}^2) + \frac{\Re\{Q_1(Z)\}}{(u+x) \sqrt{1 - (u-x)^{2}}}
\end{equation}
and
\begin{equation}
I_2 = (u p_x + x p_u)^{2} - \tfrac14 (p_u - p_x)^{2}+
      \frac{\Re\bigl\{\bigl[u - x +i \sqrt{1 - (u-x)^{2}} \,\bigr]
             Q_{1}(Z)\bigr\}}
           {(u+x) \sqrt{1 - (u-x)^{2}}} \ .
\end{equation}

%----------------------------------------------------------------------------%
\subsection{The polar-elliptical case of the second kind ($3 \hat 5$)}

The polar-elliptical case ($3 \hat 5$) admits only standard separation. the conformal coordinate transformation is (with $\hat \Delta = 1$)
\begin{equation}
 \begin{split}
 &\Sigma_3 (\zeta) = \zeta^{2} \ ,\quad 
   \hat\Sigma_5(\hat\zeta) = \hat\zeta^{2} + 1 \\[3pt]
 &W = A_3(\zeta) =  \ln{|\zeta|} \ ,\quad F_3( W) = e^{ W} \\[3pt]
 & \hat W = \hat A_5(\hat\zeta) = { \rm arcsinh} \hat\zeta \ ,\quad
    \hat F_5(\hat W) = \sinh{\hat  W}
 \end{split}
\end{equation}
or in non-null coordinates
\begin{equation}
 \begin{split}
   U &= \tfrac12 \Bigl[ \ln(u+x)
         + \ln\bigl(u-x+\sqrt{(u-x)^{2}+1}\,\bigr)\Bigr] \\[5pt]
   X &= \tfrac12 \Big[ \ln\bigl(u+x)
         - \ln(u-x+\sqrt{(u-x)^{2}+1}\,\bigr)\Bigr] \ .
 \end{split}  
\end{equation}
The conformal factor is 
\begin{equation}
\Psi = e^{W} \cosh{ \hat  W} = (u+x) \sqrt{(u-x)^{2}+1}\end{equation} 
so that
\begin{equation}
\Psi_{1} = \tfrac12 e^{2 U} \ , \quad \Psi_{2} = \tfrac12 e^{2 X} \ .\end{equation} 
This leads to a physical potential of the form
\begin{equation}\label{SEPP35}
\Phi = \frac{f_{1}(U)+f_{2}(X)}{e^{2U}+e^{2 X}} \ .
\end{equation} 
The Hamiltonian and the second invariant are given by
\begin{equation}
   H =  \tfrac12(-p_u{}^2+p_x{}^2) + 
   \frac{f_{1}(U)+f_{2}(X)}{(u+x) \sqrt{(u-x)^{2}+1}}
\end{equation}
and
\begin{equation}
I_2 = (u p_x + x p_u)^{2} + \tfrac14 (p_u - p_x)^{2}+ 
            \left[\frac{u-x}{\sqrt{(u-x)^2+1}}-1\right] f_1(U)
           +\left[\frac{u-x}{\sqrt{(u-x)^2+1}}+1\right] f_2(X) \ .
\end{equation}

%----------------------------------------------------------------------------%
\subsection{The elliptic-hyperbolic case of the first kind ($4 \hat 4$)}
The cases $(4 \hat 4)_{\epsilon}$ represent the first kind of combination of hyperbolic and elliptical coordinates and admit standard and harmonic separation: the conformal coordinate transformation is (with $\Delta = \hat \Delta = 1$)
\begin{equation}
 \begin{split}
 &\Sigma_4 (\zeta) = \zeta^{2}-1 ,\quad \hat\Sigma_4(\hat\zeta) = \hat\zeta^{2} - 1 \\[3pt]
 & W = A_4(\zeta) =  \int \frac{d \zeta}{\sqrt{\zeta^{2}-1}} = \begin{cases}
  { \rm arcosh}\zeta\ , & F_4(W) = \cosh W\ , \quad |u+x|>1 \\
    \arcsin{\zeta}\ , & F_4(W) = \sin W\ , \quad  |u+x|<1
     \end{cases}  \\[5pt]
 & \hat W = \hat A_4(\hat\zeta) = \int \frac{d \hat\zeta}{\sqrt{\hat\zeta^{2}-1}} = \begin{cases}
   { \rm arcosh}\hat\zeta\ , & \hat F_4(\hat W) = \cosh\hat W\ , \quad  |u-x|>1 \\
    \arcsin\hat\zeta\ , & \hat F_4(\hat W) = \sin\hat W\ , \quad |u-x|<1 
     \end{cases}  
 \end{split}
\end{equation}
so that the two occurrences are given by
\begin{equation}
 \begin{split}
& \epsilon = +1\ , \quad u^{2}-x^{2}>1 \quad \text{or} \quad u^{2}-x^{2}<1\ , \quad \text{H-J separation} \\
& \epsilon = -1\ , \quad |u+x|<1 \quad \text{or} \quad |u-x|<1\ ,  \quad \text{harmonic separation}  \ .
\end{split}
\end{equation}
In non-null coordinates, in the standard Hamilton-Jacobi case, we have
\begin{equation}
 \begin{split}
   U &= \tfrac12 \bigl[ { \rm arcosh}(u+x) +  { \rm arcosh}(u-x) \bigr]\ ,
   \quad X = \tfrac12 \bigl[ { \rm arcosh}(u+x) -  { \rm arcosh}(u-x) \bigr]\ , \quad u^{2}-x^{2}>1 \\[5pt]
   U &= \tfrac12 \bigl[ \arcsin(u+x) + \arcsin(u-x) \bigr]\ ,
   \quad X = \tfrac12 \bigl[ \arcsin(u+x) - \arcsin(u-x) \bigr]\ , \quad u^{2}-x^{2}<1
\end{split} 
\end{equation}
whereas, in the harmonic case, we have
\begin{equation}
   U = \tfrac12 \bigl[  { \rm arcosh}(u+x) + \arcsin(u-x) \bigr]\ ,
   \quad X = \tfrac12 \bigl[  { \rm arcosh}(u+x) - {\arcsin}(u-x) \bigr] \ ,  
   \end{equation}
for $ |u-x|<1 $ and the same, but  with $x \rightarrow - x$, for $|u+x|<1$.

%----------------------------------------------------------------------------%
\subsubsection{Case $(4\hat4)_+$}
The conformal factor for this Hamilton-Jacobi case is 
\begin{equation}\label{psi44+}
\Psi = \sinh{W} \sinh{ \hat  W} = 
\sqrt{\bigl[(u-x)^{2}-1\bigr]\bigl[(u+x)^{2}-1\bigr]}
\end{equation} 
so that
\begin{equation}
\Psi_{1} = \tfrac12 \cosh{2 U} \ , \quad \Psi_{2} = -\tfrac12 \cosh{2 X} \ .\end{equation} 
This leads to a physical potential of the form
\begin{equation}\label{SEPP44+}
\Phi = \frac{f_{1}(U)+f_{2}(X)}{\cosh{2 U}-\cosh{2 X}} \ .
\end{equation} 
The Hamiltonian and the second invariant are given by
\begin{equation}
   H =  \tfrac12(-p_u{}^2+p_x{}^2) + 
   \frac{f_{1}(U)+f_{2}(X)}
        {\sqrt{\bigl[(u-x)^{2}-1\bigr]\bigl[(u+x)^{2}-1\bigr]}}
\end{equation}
and
\begin{equation}
I_2 = (u p_x + x p_u)^{2} - \tfrac12 (p_u{}^2+p_x{}^2)+ 
  \frac{1}{\Psi}
  {\Bigl[(u^{2}-x^{2}-\Psi ) f_1(U) + (u^{2}-x^{2}+\Psi ) f_2(X) \Bigr]}
\end{equation}
with $\Psi$ given by (\ref{psi44+}).

%----------------------------------------------------------------------------%
\subsubsection{Case $(4\hat4)_-$}
In the region defined by $|u-x|<1$, the separating variables for this harmonic case is given by
\begin{equation}
 \begin{split}
   Z&=\tfrac12(1+i)\bigl[ { \rm arcosh}(u+x) +i\arcsin(x-u)\bigr] \\[4pt]
       \bar Z&=\tfrac12(1-i)\bigl[ { \rm arcosh}(u+x) -i\arcsin(x-u)\bigr]
 \end{split}
\end{equation}
with the conformal factor
\begin{equation}
\Psi = \sinh{W} \cos{\hat  W}
     = \sqrt{\bigl[1 - (u-x)^{2}\bigr]\bigl[(u+x)^{2}-1\bigr]}
     = \tfrac12 \bigl[\sinh\bigl({(1-i)Z}\bigr)
        + \sinh\bigl({(1+i){\bar Z}}\bigr)\bigr] \ .
\end{equation}
This leads to a physical potential of the form
\begin{equation}\label{SEPP44-}
\Phi = \frac{\Re\Bigl\{Q_1\Bigl(\tfrac12(1+i)
                 \bigl[ { \rm arcosh}(u+x) +i\arcsin(x-u)\bigr] \Bigr) \Bigr\}}
                 {\sqrt{\bigl[1 - (u-x)^{2}\bigr]\bigl[(u+x)^{2}-1\bigr]}} \ ,\end{equation} 
where $Q_{1}$ is an arbitrary function. The Hamiltonian and the second invariant are given by
\begin{equation}
   H = \tfrac12(-p_u{}^2+p_x{}^2) + \frac{\Re\{Q_{1}(Z)\}}
            {\sqrt{\bigl[1 - (u-x)^{2}\bigr]\bigl[(u+x)^{2}-1\bigr]}}
\end{equation}
and
\begin{equation}
I_2 = (u p_x + x p_u)^{2} - \tfrac12 (p_u{}^2+p_x{}^2)
            - \frac{\Re\bigl\{\sin\bigl((1-i){\bar Z}\bigr) Q_{1}(Z)\bigr\}}
{\sqrt{\bigl[1 - (u-x)^{2}\bigr]\bigl[(u+x)^{2}-1\bigr]}} \ .
\end{equation}
In the region $|u+x|<1$ we can recover the harmonic separation for this case by the transformation $x \rightarrow - x$.

%----------------------------------------------------------------------------%
\subsection{The elliptic-hyperbolic case of the second kind ($4 \hat 5$)}

The cases $(4 \hat 5)_{\epsilon}$ represent the second kind of combination of hyperbolic and elliptical coordinates and admit standard and harmonic separation. The conformal coordinate transformation is (with $\Delta = \hat \Delta = 1$)
\begin{equation}
 \begin{split}
 &\Sigma_4 (\zeta) = \zeta^{2}  - 1 \ ,\quad \hat\Sigma_5(\hat\zeta)
  = \hat\zeta^{2} + 1 \\[5pt]
 & W = A_4(\zeta) = \int \frac{d \zeta}{\sqrt{\epsilon(\zeta^{2}-1)}}
  = \begin{cases}
   {\rm arccosh} \zeta \ , & F_4(W) = \cosh{W}\ , \quad \epsilon=+1 \ , 
    \quad \text{H-J separation} \\[3pt]
    \arcsin{\zeta} \ , & F_4(W) = \sin{W}\ , \quad \epsilon=-1 \ , 
    \quad \text{harmonic separation}    \end{cases}  \\[5pt]
 &  \hat W = \hat A_5(\hat\zeta) ={\rm arcsinh}\hat\zeta \ ,\quad
    \hat F_5(\hat W) = \sinh{\hat  W}
 \end{split}
\end{equation}
or in non-null coordinates
\begin{equation}
 \left.
 \begin{split}
   U &= \tfrac12 \bigl[ { \rm arcosh}(u+x) +  {\rm arcsinh}(u-x) \bigr] \\[3pt]
   X &= \tfrac12 \bigl[ { \rm arcosh}(u+x) -  {\rm arcsinh}(u-x) \bigr]
 \end{split} \ \right\} \quad \epsilon=+1 \ ,\quad |u+x| > 1
\end{equation}
and
\begin{equation}
 \left.
 \begin{split}
   U &= \tfrac12 \bigl[ \arcsin(u+x) +  {\rm arcsinh}(u-x)\bigr] \\[3pt]
   X &= \tfrac12 \bigl[ \arcsin(u+x) -  {\rm arcsinh}(u-x)\bigr]
 \end{split} \ \right\} \quad \epsilon=-1 \ , \quad |u+x| < 1 \ .
\end{equation}

%----------------------------------------------------------------------------%
\subsubsection{Case $(4\hat5)_+$}
The conformal factor is 
\begin{equation}\label{psi45+}
\Psi = \sinh{W} \cosh{ \hat  W} = 
\sqrt{\bigl[(u+x)^{2}-1\bigr]\bigl[(u-x)^{2}+1\bigr]}
\end{equation} 
so that
\begin{equation}
\Psi_{1} = \tfrac12 \sinh{2 U} \ , \quad \Psi_{2} = \tfrac12 \sinh{2 X} \ .\end{equation} 
This leads to a physical potential of the form
\begin{equation}
\label{SEPP45+}
\Phi = \frac{f_{1}(U)+f_{2}(X)}{\sinh{2 U}+ \sinh{2 X}} \ .
\end{equation} 
The Hamiltonian and the second invariant are given by
\begin{equation}
   H =  \tfrac12(-p_u{}^2+p_x{}^2) + 
   \frac{f_{1}(U)+f_{2}(X)}{\sqrt{\bigl[(u+x)^{2}-1\bigr]
                              \bigl[(u-x)^{2}+1\bigr]}}
\end{equation}
and
\begin{equation}
I_2 = (u p_x + x p_u)^{2} - p_u p_x + 
  \frac{1}{\Psi}
  {\Bigl[(u^{2}-x^{2}-\Psi ) f_1(U) + (u^{2}-x^{2}+\Psi ) f_2(X) \Bigr]}
\end{equation}
with $\Psi$ taken from (\ref{psi45+}).

%----------------------------------------------------------------------------%
\subsubsection{Case $(4\hat5)_-$}
In the harmonic case, the separating variables are
\begin{equation}
 \begin{split}
 Z&=\tfrac12(1+i)\bigl[ {\arcsin}(u+x) +i{\rm arcsinh}(x-u)\bigr] \\[3pt]
 \bar Z&=\tfrac12(1-i)\bigl[\arcsin(u+x) -i{\rm arcsinh}(x-u)\bigr]
 \end{split}
\end{equation}
with the conformal factor
\begin{equation}
\Psi = \sinh{W} \cos{\hat  W}
     = \sqrt{\bigl[1 + (u-x)^{2}\bigr]\bigl[1-(u+x)^{2}\bigr]}
\end{equation}
which can be written in the form
\begin{equation}
\Psi = \tfrac12 \bigl[\cos\bigl({(1-i)Z}\bigr)
                 + \cos\bigl({(1+i){\bar Z}}\bigr)\bigr] \ .
\end{equation}
This leads to a physical potential of the form
\begin{equation}\label{SEPP45-}
\Phi = \frac{\Re\Bigl\{Q_1\Bigl(
           \tfrac12(1+i)
              \bigl[\arcsin(u+x) +i{\rm arcsinh}(x-u)\bigr] \Bigr) \Bigr\}}
                 {\sqrt{\bigl[1 + (u-x)^2\bigr]\bigl[1-(u+x)^2\bigr]}} \ ,
\end{equation} 
where $Q_{1}$ is an arbitrary function. The Hamiltonian and the second invariant are given by
\begin{equation}
   H = \tfrac12(-p_u{}^2+p_x{}^2)
       + \frac{\Re\{Q_{1}(Z)\}}
              {\sqrt{\bigl[1 + (u-x)^{2}\bigr]\bigl[1-(u+x)^{2}\bigr]}}
\end{equation}
and
\begin{equation}
I_2 =  (u p_x + x p_u)^{2} - p_u p_x
      + \frac{\Im\bigl\{\cos\bigl((1+i){\bar Z}\bigr) Q_{1}(Z)\bigr\}}
        {\sqrt{\bigl[1 + (u-x)^{2}\bigr]\bigl[1-(u+x)^2\bigr]}} \ .
\end{equation}

%----------------------------------------------------------------------------%
\subsection{The elliptic-hyperbolic case of the third kind ($5 \hat 5$)}

The elliptic-hyperbolic case ($5 \hat 5$) admits only Hamilton-Jacobi separation. The conformal coordinate transformation is (with $\Delta = \hat \Delta = 1$)
\begin{equation}
 \begin{split}
 &\Sigma_5 (\zeta) = \zeta^{2}+1\ ,\quad \hat\Sigma_5(\hat\zeta)
                   = \hat\zeta^{2} + 1 \\[3pt]
 &W = A_5(\zeta) ={\rm arcsinh}\zeta \ , \quad F_5( W) = \sinh{ W} \\[3pt]
 & \hat W = \hat A_5(\hat\zeta)
          ={\rm arcsinh}\hat\zeta \ ,\quad  \hat F_5(\hat W) = \sinh{\hat  W}
 \end{split}
\end{equation}
or in non-null coordinates
\begin{equation}
 \begin{split}
   U &= \tfrac12 \bigl[{\rm arcsinh}(u+x) +{\rm arcsinh}(u-x)\bigr] \\[3pt]
   X &= \tfrac12 \bigl[{\rm arcsinh}(u+x) -{\rm arcsinh}(u-x)\bigr] \ .
 \end{split}  
\end{equation}
The conformal factor is
\begin{equation}\label{psi55}
\Psi = \cosh{W} \cosh{ \hat  W} = 
\sqrt{\bigl[(u-x)^{2}+1\bigr]\bigl[(u+x)^{2}+1\bigr]}
\end{equation} 
so that
\begin{equation}
\Psi_{1} = \tfrac12 \cosh{2 U} \ , \quad \Psi_{2} = \tfrac12 \cosh{2 X} \ .\end{equation} 
This leads to a physical potential of the form
\begin{equation}
\label{SEPP55}
\Phi = \frac{f_{1}(U)+f_{2}(X)}{\cosh{2 U}+\cosh{2 X}} \ .
\end{equation} 
The Hamiltonian and the second invariant are given by
\begin{equation}
   H =  \tfrac12(-p_u{}^2+p_x{}^2) + 
   \frac{f_{1}(U)+f_{2}(X)}
        {\sqrt{\bigl[(u-x)^{2}+1\bigr]\bigl[(u+x)^{2}+1\bigr]}}
\end{equation}
and
\begin{equation}
I_2 = (u p_x - x p_u)^{2} + \tfrac12 (p_u{}^2+p_x{}^2)+ 
  \frac{1}{\Psi}
  {\Bigl[(u^{2}-x^{2}-\Psi ) f_1(U) + (u^{2}-x^{2}+\Psi ) f_2(X) \Bigr]}
\end{equation}
with $\Psi $ as in (\ref{psi55}).

%----------------------------------------------------------------------------%
\section{Comments and conclusions}
The main result presented in this paper is the generalization of the standard separability notion for $(1+1)$-dimensional systems to include also the complex/harmonic and the linear/null separation structures. These new separability structures appear naturally in the approach to separability based on the standardization of the traceless conformal Killing tensor by means of suitable conformal coordinate transformations. In the case of a metric with indefinite signature, the determinant of the conformal Killing tensor may vanish and change sign in contrast to the case of positive definite metrics in which the determinant is always positive. The new separability structures arise precisely when the determinant is negative or zero. The analytical justification for this phenomenon relies on the forms taken by the generalized Darboux equation in the three cases: in the standard Hamilton-Jacobi case, it has the same form (wave equation) as its counterpart for positive definite systems. In the complex/harmonic case on the other hand, it takes the form of the Laplace equation and the separating coordinates are complex conjugate pairs. In the linear/null case, the generalized Darboux equation depends only on one of the separating variables; the other coordinate can be chosen freely.

In the usual case of separability for arbitrary values of the Hamiltonian function (the strong integrability case which includes also free  motion on the flat hyperbolic plane) we have listed all separating coordinate systems, including those associated with these new separability structures. The complex/harmonic separation appears in five cases: the Cartesian-Cartesian $(1\hat1)_-$ case; the Cartesian-parabolic $(1\hat2)_-$ case; the parabolic-parabolic $(2\hat2)_-$ case; the elliptic-hyperbolic cases of first $(4\hat4)_-$ and second $(4\hat5)_-$ kind. The linear/null separation appears in three cases: the Cartesian-Cartesian $(1\hat1)_0$ case; the Cartesian-parabolic $(1\hat2)_0$ case; the parabolic-parabolic $(2\hat2)_0$ case. 

There are several directions in which one can extend and generalize this work. We mention two of them. The first is that of exploring the issue of the weak integrability: in the indefinite case, Hamiltonian systems integrable on a single energy surface only are more interesting than those in the positive definite case, because they can be used to describe systems with time-dependent potentials \cite{db,mb}. A second issue is connected with integrable but nonseparable systems: in analogy with the results obtained in the positive definite case \cite{max,kprs}, it is quite natural to assume the existence of systems on the Minkowski plane admitting integrals of motion cubic or quartic in the momenta.

\section*{Acknowledgment}

We wish to thank the referee for his extensive and useful  comments  which have 
prompted us to make an attempt to improve the  presentation.


\begin{thebibliography}{10}
\bibitem{nota1}
In our treatment we have chosen notation in such a way that every quantity expressed by a Latin letter in the positive definite case has its indefinite counterpart expressed by the corresponding Greek letter.
\bibitem{nota2}
  In the following, $f_1$ and $f_2$ will always stand for arbitrary functions of their arguments.
\end{thebibliography}
\end{document}